\documentclass[12pt]{article}
\pdfoutput=1
\usepackage[pdftex]{graphicx,color}
\usepackage{epsfig}
\usepackage{latexsym,amsmath,amsfonts,amssymb}
\usepackage[latin1]{inputenc}
\usepackage[american]{babel}
\usepackage{hyperref}
\usepackage{bbm}
\begin{document}

\newcommand{\ket}[2]{\langle #1,#2\rangle}
\newcommand{\bra}[2]{\left[ #1,#2\right]}
\newcommand{\Pgen}[3]{\langle#1\lvert#2\lvert#3]}

\thispagestyle{empty}
\begin{center}
\Large{\textbf{The Twelve-Graviton Next-to-MHV Amplitude from Risager's Construction}} \\
\vspace{10mm}
\large\text{Eduardo Conde$^1$ and Sayeh Rajabi$^{2,3}$}\\
\vspace{15mm}
\normalsize \textit{$^1$Departamento de  F\'\i sica de Part\'\i  culas, Universidade de Santiago de Compostela\\
and\\
Instituto Galego de F\'\i sica de Altas Enerx\'\i as (IGFAE)\\
E-15782, Santiago de Compostela, Spain} \\
\vspace{4mm}
\normalsize \textit{$^2$Perimeter Institute for Theoretical Physics, Waterloo, ON, N2L 2Y5, CA} \\
\vspace{4mm}
\normalsize\textit{$^3$Department of Physics and Astronomy $\&$ Guelph-Waterloo Physics Institute,} \\
\normalsize \textit{University of Waterloo, Waterloo, ON, N2L 3G1, CA}\\ 

\let\thefootnote\relax\footnotetext{ eduardo@fpaxp1.usc.es, srajabi@perimeterinstitute.ca}
\end{center}
\vspace{20mm}

\abstract

The MHV or CSW expansion of tree-level Yang-Mills amplitudes provides an elegant and simple way of obtaining analytic formulas for S-matrix elements. Inspired by the BCFW technique, a systematic approach to obtain the MHV expansion was introduced by Risager, using a particular complex deformation. Although it works for Yang-Mills amplitudes, Risager's technique fails to provide an MHV expansion already for Next-to-MHV gravity amplitudes with more than eleven particles, as shown by Bianchi, Elvang and Freedman in 2008~\cite{Bianchi:2008pu}. This fact implies that in this sector there is a contribution at infinity starting at $n = 12$. In this note we determine the explicit analytic form of this residue at infinity for $n = 12$. Together with the terms of the Risager MHV expansion, the residue at infinity completes the first full CSW-like analytic expression of the twelve-graviton NMHV amplitude. Our technique can also be used to compute the residue at infinity for higher points.
\newpage

\section{Introduction and Summary}
\setcounter{page}{1}

Tree-level gravity amplitudes are objects of genuine theoretical interest. Although in practice they can be constructed with Britto-Cachazo-Feng-Witten recursion relations \cite{BCFW,taming,Nima Jared}, it is of great interest to have analytic formulas for them (see for instance~\cite{Berends:1988zp, Bedford:2005yy, Elvang:2007sg, Mason:2008jy, Spradlin:2008bu, Drummond:2009ge, Nguyen:2009jk}). In particular, the simpler these formulas are, the more insight they contain about tree-level gravity. A great step towards this goal has been recently taken by Hodges, who found an extremely elegant formula for Maximally-Helicity-Violating gravity amplitudes \cite{Hodges:2012ym}. This renews the interest in developing an MHV-vertex expansion for gravity amplitudes.

In 2005 Risager \cite{Risager} proposed a certain complex deformation which, by an induction procedure, proved that Yang-Mills tree-level amplitudes can be constructed using the Cachazo-Svrcek-Witten rules \cite{CSW}. Applying Risager's technique to gravity amplitudes was the next natural goal \cite{BjerrumBohr:2005jr}. However, as verified by numerical calculations in \cite{Bianchi:2008pu} and later by analytic means in \cite{taming}, graviton amplitudes in the Next-to-MHV sector depend on the reference spinor of Risager deformation, starting at twelve particles. 

In the present note, we address the question of how the Risager expansion disagrees with the physical amplitude in the NMHV sector of gravity, \textit{i.e.} we study the tree-level amplitude $M_n\left(1^-,2^-,3^-,4^+,\ldots,n^+\right)$ for $n\geq12$, and develop a procedure to determine this discrepancy. As an illustration, we present the explicit result in the case of the twelve-particle amplitude.

Let us specify our notation. We denote the NMHV amplitude of our interest simply by $M_n$. We use the spinor-helicity formalism and represent the momenta of the gravitons as $p_i=\lambda^{(i)}\tilde\lambda^{(i)}$, $(i=1,\ldots,12)$. The Risager shift deforms the anti-holomorphic spinors of the three negative-helicity particles as
\begin{equation}
     \left\{ \begin{array}{rcc}
          \tilde\lambda^{(1)}(w)=\tilde\lambda^{(1)}+w\,\ket{2}{3}X \\
           \tilde\lambda^{(2)}(w)=\tilde\lambda^{(2)}+w\,\ket{3}{1}X \\
           \tilde\lambda^{(3)}(w)=\tilde\lambda^{(3)}+w\,\ket{1}{2}X 
        \end{array} \right. \,; \quad\quad\quad  M_n\rightarrow M_n(w)\,,     
         \label{eqn:def.Ris} 
\end{equation}
where X is an arbitrary reference spinor, and $w$ is the complex variable that we associate with the Risager shift (we later associate $z$ with BCFW shifts). We have then a one-parameter family of amplitudes $M_n(w)$. We call $M_n=M_n(0)$ the \textit{physical amplitude}, for obvious reasons, while we denote by \textit{Risager expansion}, $M_{n}^{\textrm{Ris}}$, the sum of residues of $M_n(w)$ at its poles on the complex plane. The Risager expansion can be expressed as the following MHV-vertex decomposition: 
\begin{equation}
M_{n}^{\textrm{Ris}}=\sum_{a,L^{+}}M_{n_L}\left(\hat a^{-},L^{+},\left(-I\right)^{-}\right)\frac{1}{P_L^2}\,M_{n_R}\left(I^{+},\hat b^{-},\hat c^{-},R^{+}\right)\,,
\label{eqn:Risexp}
\end{equation}
where by $P_L$ we mean $P_L=p_a+P_{L^+}=p_a+\sum_{l_i\in L^{+}}p_{l_i}$, and the labels $a,b,c$ denote negative-helicity gravitons, whereas $l,l_1,l_2,\ldots,r_1,r_2,\ldots$ denote positive-helicity ones. $L^{+}$ ($R^{+}$) denotes the subset of external positive-helicity gravitons in the left (right) sub-amplitudes in \eqref{eqn:Risexp}. We use $n_L$ ($n_R$) for the number of external legs in the left (right), so that $n+2=n_L+n_R$. The hats on particles $1,2,3$ indicate that, in each of the sub-amplitudes of \eqref{eqn:Risexp}, their momenta must be evaluated with \eqref{eqn:def.Ris} at the appropriate value of $w=\hat{w}$ (the one that makes $(p_{\hat{a}}+P_{L^{+}})^2=0$). The momentum of the graviton $I$ (opposite to the momentum of graviton $-I$) is determined by momentum conservation.

It is known \cite{Bianchi:2008pu} (see also appendix B of \cite{taming}) that the Risager deformation fails to give a valid recursion relation for $n\geq12$, since $M_n(w)\sim w^{n-12}$ as $w\to\infty$. In order to fix the Risager expansion for $n\geq 12$, one needs to compute the residue at infinity, that we denote by ${\cal A}_n$, which can be defined as
\begin{equation}
{\cal A}_{n}=M_{n}-M_{n}^{\textrm{Ris}}\,.
\label{eqn:def.a}
\end{equation}
Of course, we have that ${\cal A}_{n}=0$ for $n<12$. 

Our method for computing ${\cal A}_{n}$ is as follows: we perform a BCFW complex deformation on two external legs, making ${\cal A}_n\to{\cal A}_n(z)$, that allows us to recover the original object ${\cal A}_n$ from the residues at its poles. This can be done since under certain BCFW deformations, $M_n^{\text{Ris}}(z)\rightarrow 0$ at large $z$, as we discuss in section~\ref{sec:bcfwcomp}. It happens that the $z$-dependent poles of ${\cal A}_{n}(z)$ can be split into physical and unphysical ones. The physical poles are, as usual, of the form $1/P^2(z)$ where $P(z)$ is the sum of external momenta in one sub-amplitude. The unphysical poles depend on the reference spinor $X$. The result for the $n$-point residue at infinity is then
\begin{equation}
{\cal A}_n=-\sum_{\textrm{phys}}\textrm{Res}\left[\frac{{\cal A}_n(z)}{z}\right]-\sum_{\textrm{unphys}}\textrm{Res}\left[\frac{{\cal A}_n(z)}{z}\right]\,.
\label{eqn:sum.a12}
\end{equation}
We explicitly calculate ${\cal A}_{12}$ (performing the BCFW deformation on particles 1 and 4), and get
\begin{equation}\label{eqn:unphys}\begin{aligned}
&\sum_{\textrm{unphys}}\textrm{Res}\left[\frac{{\cal A}_{12}}{z}\right]=-\left(\ket{1}{2}\ket{2}{3}\ket{3}{1}\right)^6\prod_{k=5}^{12}\frac{\bra{k}{X}}{\ket{1}{k}\ket{2}{k}\ket{3}{k}}\\
&\times\sum_{l=5}^{12}\frac{\bra{4}{l}^6}{\bra{4}{X}^2\bra{l}{X}^2}\frac{\ket{4}{l}\bra{4}{l}}{\Pgen{1}{p_4+p_l}{X}\Pgen{2}{p_4+p_l}{X}\Pgen{3}{p_4+p_l}{X}}\frac{\ket{1}{l}\ket{2}{l}\ket{3}{l}}{\bra{l}{X}}\,,
\end{aligned}\end{equation}
for the sum of residues at the unphysical poles (we have used the standard notation $\Pgen{i}{\sum_j p_j}{X}=\sum_j\ket{i}{j}\bra{j}{X}$), and at the physical poles we have:
\begin{equation}\label{eqn:phys}\begin{aligned}
\sum_{\textrm{phys}}\textrm{Res}\left[\frac{{\cal A}_{12}}{z}\right]=&-\ket{3}{1}\frac{\bra{4}{X}\ket{1}{2}^2}{\ket{2}{4}\ket{1}{4}^2}\textrm{Res}\left[M_{11}^{A}(w),\infty\right]\\
&-\ket{1}{2}\frac{\bra{4}{X}\ket{1}{3}^2}{\ket{3}{4}\ket{1}{4}^2}\textrm{Res}\left[M_{11}^{B}(w),\infty\right]\,,\
\end{aligned}\end{equation}
where $M_{11}^{A}$ and $M_{11}^{B}$ are eleven-point NMHV amplitudes that are obtained by ``dissolving'' particle 4 into particles $(1,2)$ and $(1,3)$ respectively, and one performs the Risager shift \eqref{eqn:def.Ris} on them to obtain $M_{11}^{A}(w)$ and $M_{11}^{B}(w)$. The precise meaning of ``dissolving'' is defined in section~\ref{sec:12anomaly}.

As usual with formulas obtained from BCFW, \eqref{eqn:unphys} and \eqref{eqn:phys} are asymmetric in the set of positive helicity particles (note that the deformed particle 4 is special in the formulas), but the sum of them is indeed invariant under permutation of the positive labels.

\section{The Residue at Infinity}

The behavior of the deformed amplitude $M_n(w)$ at infinity, $M_n(w)\sim 1/w^{12-n}$, makes it clear that starting at $n=12$ particles, there is a contribution at infinity, ${\cal A}_n$, that must be added to the Risager expansion in order to recover the physical amplitude. However, it is useful to have an alternative perspective on the existence of this residue at infinity, namely a more physical reason why it appears. Notice that it could be that this contribution at infinity vanished for $n>12$. We show that this is not the case.

For BCFW two-particle deformations, the presence of a contribution at infinity is related to the BCFW amplitude missing some physical factorization channels \cite{Benincasa:2011kn}. For the Risager three-particle deformation, the situation is different. The Risager expansion does not miss any physical pole; rather, it contains extra residues as well as some unphysical poles. Let us be more specific.

\subsection{Physical Meaning of the Residue at Infinity}
\label{ssec:phys}

One can ask about the factorization channels of the NMHV scattering amplitude of $n$ gravitons that are correctly reproduced by the Risager expansion $M_n^{\textrm{Ris}}$. In view of the definition \eqref{eqn:def.a}, one can also search for the poles of ${\cal A}_n$.

It turns out that all the poles of the physical amplitude $M_n$ are already present in $M_n^{\textrm{Ris}}$. Moreover, most of the residues of $M_n^{\textrm{Ris}}$ at these poles give the expected factorization of the physical amplitude $M_n$. It happens that the physical factorization fails in two classes of channels: the ones corresponding to the poles $\ket{l_1}{l_2}$ and $\ket{a}{l}$ (recall $l,l_1,l_2$ denote positive-helicity gravitons and $a$ is a negative-helicity graviton).

We first study the limit $\ket{l_1}{l_2}\to0$. In this limit, the only singular diagrams in the expansion \eqref{eqn:Risexp} of $M_n^{\textrm{Ris}}$ are the ones that have both gravitons $l_1,l_2$ either on the left or on the right sub-amplitude. It is then easy to see that we have the following factorization:
\begin{equation}
\lim_{\ket{l_1}{l_2}\to0}\ket{l_1}{l_2}\bra{l_1}{l_2}M_{n}^{\textrm{Ris}}=M_3\left(l_1^+,l_2^+,p_{l_1l_2}^-\right)M_{n-1}^{\textrm{Ris}}\,,
\label{eqn:l1l2}
\end{equation}
where it is understood that in the $(n-1)$-point Risager expansion $l_1$ and $l_2$ are substituted by a positive-helicity graviton with on-shell momentum $p_{l_1l_2}=p_{l_1}+p_{l_2}$. Equation \eqref{eqn:l1l2} implies that the residue at this type of pole is the physical one as long as $M_{n-1}^{\textrm{Ris}}=M_{n-1}$, which holds for $n<13$.

The limit $\ket{a}{l}\to0$ is a little bit more subtle. The singular diagrams of $M_n^{\textrm{Ris}}$ are now those where particles $\hat{a}$ and $l$ are on the same sub-amplitude, and a three-particle amplitude $M_3\left(\hat{a}^{-},l^{+},-J^{+}\right)$ factorizes out since $p_J=p_{\hat{a}}+p_l$ becomes on-shell. The subtlety arises because $p_{\hat{a}}\neq p_a$, and such a three-point amplitude is not common to all the diagrams in the Risager expansion. Taking into account that
\begin{equation}
M_3\left(\hat{a}^{-},l^{+},-J^{+}\right)=\frac{\bra{\hat{a}}{l}^2}{\bra{a}{l}^2}\,M_3\left(a^{-},l^{+},-p_{al}^{+}\right)\,,
\end{equation}
we can write (schematically)
\begin{equation}
\lim_{\ket{a}{l}\to 0}\ket{a}{l}\bra{a}{l}\, M_{n}^{\textrm{Ris}}=M_3\left(a^{-},l^{+},-p_{al}^{+}\right)\sum\frac{\bra{\hat{a}}{l}}{\bra{a}{l}}\left(
\begin{array}{c}
\textrm{term in the Risager}\\
\textrm{expansion of }M_{n-1}
\end{array}\right)\,,
\end{equation}
where the sum is over the terms of the Risager expansion of an $(n-1)$-point amplitude $M_{n-1}$ with the same external states as $M_n$, but where $a$ and $l$ combine into a negative-helicity graviton with momentum $p_{al}=p_{a}+p_l$. Now, computing
\begin{equation}
\frac{\bra{\hat{a}}{l}}{\bra{a}{l}}-1=\hat{w}\,\frac{\bra{X}{l}}{\bra{a}{l}}\ket{b}{c}\,,
\end{equation}
it is straightforward to see that
\begin{equation}
\lim_{\ket{a}{l}\to 0}\ket{a}{l}\bra{a}{l}\, M_{n}^{\textrm{Ris}}=M_3\left(a^{-},l^{+},-p_{al}^{+}\right)\left(M_{n-1}^{\textrm{Ris}}+\ket{b}{c}\frac{\bra{X}{l}}{\bra{a}{l}}\textrm{Res}\left[M_{n-1}(w),\infty\right]\right)\,.
\label{eqn:resaln}
\end{equation}
For our Risager deformation \eqref{eqn:def.Ris}, we recall to the reader that the residue for an $n$-point NMHV amplitude can be written as:
\begin{equation}
\textrm{Res}\left[M_{n}(w),\infty\right]=\sum_{a,L^{+}}M_{n_L}\left(\hat a^{-},L^{+},\left(-I\right)^{-}\right)\frac{1}{\ket{b}{c}\Pgen{a}{P_{L}}{X}}M_{n_R}\left(I^{+},\hat b^{-},\hat c^{-},R^{+}\right)\,,
\label{eqn:Res.infty}
\end{equation}
where the notation is as in \eqref{eqn:Risexp}. The implication of equation \eqref{eqn:resaln} is that we have the proper physical factorization at the poles $\ket{a}{l}$ when, besides the previous condition $M_{n-1}^{\textrm{Ris}}=M_{n-1}$, it also happens that $\textrm{Res}\left[M_{n-1}(w),\infty\right]=0$. For this last condition to hold, $w\,M_{n-1}(w)$ must vanish at infinity, or equivalently $M_{n-1}(w)$ must vanish faster than $1/w$, which happens only for $n<12$.

Interestingly, the necessity of the residue at infinity of lower-point amplitudes to vanish also happens when reconstructing tree-level graviton amplitudes with the BCFW technique, as noted by Toro and Schuster in \cite{Schuster:2008nh}. They saw that in order to prove that the BCFW expansion for an $n$-graviton amplitude has the correct factorization in this very  same channel $\ket{a}{l}$, $(n-1)$-graviton amplitudes need to vanish faster than $1/z$ under BCFW deformations.

\smallskip

In addition to this failure to correctly reproduce physical poles, a careful analysis of the different terms in the Risager expansion \eqref{eqn:Risexp} shows that for $n\geq12$, unphysical poles of the form $\Pgen{a}{P_{L}}{X}$ appear in $M_n^{\textrm{Ris}}$. More precisely, they appear in the denominator with the power $\Pgen{a}{P_{L}}{X}^{n-7-n_L}$.

In order to have an intuition of why $n=12$ is special, let us look at a given diagram of the Risager expansion~\eqref{eqn:Risexp} with $L^{+}=\{l_1,...,l_{n_L-2}\}$ and $R^{+}=\{r_2,...,r_{n_R-2}\}$. The contribution of this diagram to the Risager expansion is
\begin{equation}
M_{n_L}\left(\hat a^{-},L^{+},\left(-I\right)^{-}\right)\frac{1}{P_L^2}\,M_{n_R}\left(I^{+},\hat b^{-},\hat c^{-},R^{+}\right)\,,
\label{eqn:Risexp1}
\end{equation}
where $M_{n_L}$ and $M_{n_R}$ are MHV amplitudes. These amplitudes could contain poles of the form $\ket{a}{I}=\Pgen{a}{P_{L}}{X}$, where by convention we use $\lambda^I=P_L|X]=\bra{a}{X}\lambda^{a}+\sum_{l_i\in L^+}\bra{i}{X}\lambda^{l_i}$. To check this possibility we can use any explicit analytic expression of MHV amplitudes. We do it using the Mason-Skinner formula \cite{Mason:2008jy}, which reads for the MHV amplitude $M_n\left(1^-,2^+,\ldots,(n-1)^+,n^-\right)$ as
\begin{equation}\begin{aligned}
&M_n^{\textrm{MHV}}=\frac{\ket{1}{n}^8}{\ket{1}{n-1}\ket{n-1}{n}\ket{n}{1}}\left(\frac{1}{\ket{1}{2}\ket{2}{3}\cdots\ket{n-1}{n}\ket{n}{1}}\right.\\
&\left.\times\prod_{k=2}^{n-2}\frac{\Pgen{n}{p_{n-1}+\ldots+p_{k+1}}{k}}{\ket{k}{n}}+\left(\textrm{permutations of labels }\{2,\ldots,n-2\}\right)\right)\\
&\qquad=\frac{\ket{1}{n}^6}{\ket{1}{n-1}\ket{n-1}{n}}\left(\frac{1}{\ket{1}{2}\ket{2}{3}\cdots\ket{n-1}{n}}\right.\\
&\left.\times\prod_{k=2}^{n-2}\frac{\Pgen{n}{-p_{1}-\ldots-p_{k-1}}{k}}{\ket{k}{n}}+\left(\textrm{permutations of labels }\{2,\ldots,n-2\}\right)\right)\,.
\end{aligned}\label{eqn:MS}\end{equation}
With our convention, $\lambda^I=P_L|X]$, the Mason-Skinner formula yields for $M_{n_L}$ a factor $\ket{a}{I}^6=\Pgen{a}{P_{L}}{X}^6$ in the numerator, since $a$ and $I$ are the negative-helicity particles on the left sub-amplitude. One can notice that the power of this factor  is the fingerprint of $\mathcal{N}=8$ SUSY (it was initially eight, before canceling two powers of the same factor in the denominator). In the expression for $M_{n_R}$, identifying $\hat{b}\equiv1$, $\hat{c}\equiv n$ and $I\equiv n-1$ when using Mason-Skinner formula~\eqref{eqn:MS}, $\ket{a}{I}$ appears only through the denominator of the complexified momentum $p_{\hat{b}}$, since with Risager deformation~\eqref{eqn:def.Ris} we evaluate the sub-amplitude at 
\begin{equation}
\hat w=-\frac{P_L^2}{\ket{b}{c}\Pgen{a}{P_{L}}{X}}\,.
\label{eqn:what}
\end{equation}
Therefore, from the product inside Mason-Skinner formula~\eqref{eqn:MS} we get $n_R-3$ powers of $\Pgen{a}{P_{L}}{X}$ in the denominator of $M_{n_R}$. In total, for the whole Risager diagram we have the power
\begin{equation}
\frac{\ket{a}{I}^6}{\ket{a}{I}^{n_R-3}}=\Pgen{a}{P_L}{X}^{9-n_R}\,.
\end{equation}
So, in order to have a pole of this type, $n_R$ needs to be at least ten. Considering that $n_L+n_R=n+2$ and $n_L\geq3$, we simply see that $n$ has to be at least eleven to produce this unphysical pole. Naively one would expect that $M_{11}^{\textrm{Ris}}$ would contain the pole\footnote{Although in this case $\Pgen{a}{P_L}{X}=\ket{a}{l}\bra{l}{X}$, notice that these diagrams with three-point amplitudes do not contribute to the poles $\ket{a}{l}$, since this factor cancels in~\eqref{eqn:what}.} $\bra{l}{X}$. However, there are three Risager diagrams contributing to this pole (the ones with $a=1,2,3$ and $L^+=\{l\}$) and, only in the case of $n=11$, a cancellation happens when summing over the three diagrams (see appendix \ref{app} for details). The pole $\bra{l}{X}$ is then spurious for $n=11$, as we knew beforehand since $M_{11}^{\textrm{Ris}}=M_{11}$. Hence, the twelve-particle amplitude is the place for the first appearance of the unphysical poles $\Pgen{a}{P_{L}}{X}$.
 
 \medskip
 
Combining all the information spelled out in this subsection, we know what the poles of ${\cal A}_{n}$ are, and we can write the schematic expression\footnote{In writing formula \eqref{eqn:an}, when $L^+=\{l\}$, by $\Pgen{a}{P_L}{X}$ we understand just $\bra{l}{X}$.}
\begin{align}
{\cal A}_{12}&=\frac{{\cal P}_{12}}{\prod\limits_{a,l,l_1,l_2}\Pgen{a}{p_{l_1}+p_{l_2}}{X}\bra{l}{X}^2\ket{a}{l}}\,,\label{eqn:a12}\\
{\cal A}_{n}&=\frac{{\cal P}_n}{\prod\limits_{\substack{a,l,l_1,l_2\\ n_L< n-7}}\Pgen{a}{P_L}{X}^{n-7-n_L}\ket{a}{l}\ket{l_1}{l_2}}\,,\qquad n>12\,,\label{eqn:an}
\end{align}
where ${\cal P}_{n}$ is some polynomial of the momenta of $n$ scattering gravitons.

\subsection{A BCFW Computation of the Residue at Infinity}
\label{sec:bcfwcomp}
\begin{figure}[h]
\centering
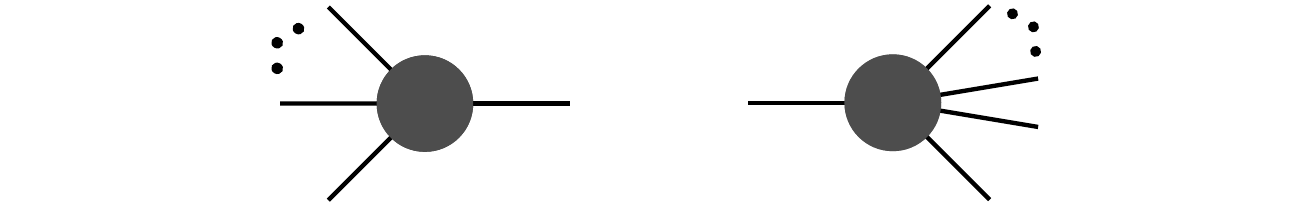
\caption{Contribution to the residue of ${\cal A}_n$ at the unphysical pole $\Pgen{a}{P_{L}}{X}$.}
\label{fig1}
\end{figure}

In virtue of \eqref{eqn:a12}-\eqref{eqn:an}, we know the poles of the contribution at infinity ${\cal A}_n$. Moreover, we also know the residues of ${\cal A}_n$ at them. At the physical poles $\ket{a}{l}$ and $\ket{l_1}{l_2}$, the residues are determined by \eqref{eqn:l1l2} and \eqref{eqn:resaln}. And at the unphysical poles $\Pgen{a}{P_{L}}{X}$, the residues come from just one diagram in the Risager expansion, namely the one with particles $a,L^+$ on the left blob (see Figure~\ref{fig1}), which is the only one that has the factor $1/\Pgen{a}{P_{L}}{X}^{n-7-n_L}$.

This information is enough to implement a one-parameter complex deformation on the momenta of some gravitons, turn the residue at infinity into a function ${\cal A}_n(z)$ of a complex variable $z$, and recover ${\cal A}_n$ from the residues at the poles of this function, as long as it vanishes at $z\rightarrow\infty$. We can actually use a BCFW shift:
\begin{equation}
\tilde\lambda^{(i)}(z)=\tilde\lambda^{(i)}-z\,\tilde\lambda^{(j)}\,,\qquad\lambda^{(j)}(z)=\lambda^{(j)}+z\,\lambda^{(i)}\,.
\label{eqn:BCFW.shift}
\end{equation}
When the helicities of particles ($i$,$j$) are respectively $(-,+),(-,-),(+,+)$, we know that $M_n$ vanishes as $1/z^2$. It is easy to check that under the last two shifts, the worst diagrams in the Risager expansion go as\footnote{Actually, it seems that the Risager expansion $M_n^{\textrm{Ris}}$ vanishes as $1/z^2$ for the shifts $(-,-),(+,+)$, exactly as the physical gravity amplitude does. We checked this numerically for $n\leq16$. The shift $(-,+)$ is not so nicely behaved, as the worst Risager diagrams go as $1/z^{13-n}$. Indeed, the Risager expansion does not vanish under the shift $(-,+)$ for $n\geq13$. For the sake of completeness, we mention that under the $(+,-)$ shift the worst Risager diagrams behave as $z^{n-5}$, and the Risager expansion displays this large-$z$ behavior for $n\geq12$.} $1/z$. From the definition \eqref{eqn:def.a}, it is obvious that ${\cal A}_n(z)$ will vanish at infinity. We can then write the usual BCFW integral:
\begin{equation}
\oint \textrm{d}z\,\frac{{\cal A}_{n}(z)}{z}=0\,\quad\implies\quad{\cal A}_{n}=-\sum_{\textrm{poles }\underline{z}}\textrm{Res}\left[\frac{{\cal A}_{n}(z)}{z};\underline{z}\right]\,,
\end{equation}
where $\underline{z}$ are the points at which some factor in the denominator of ${\cal A}_n(z)$ becomes zero. Notice that there is a difference with respect to the usual BCFW reconstruction of an amplitude. Not all the poles of ${\cal A}_n(z)$ are simple poles, since in the denominator of ${\cal A}_n$ some factors come with a higher power than one (see \eqref{eqn:an}).

Having given a procedure to compute the residue at infinity ${\cal A}_n$ for any $n$, let us illustrate it explicitly by computing the first non-zero contribution, that happens for $n=12$.

\subsection{Residue at Infinity of the Twelve-Point Amplitude}
\label{sec:12anomaly}
Following the steps outlined above, we show how to compute ${\cal A}_{12}$. Its poles were written explicitly in \eqref{eqn:a12}. The BCFW shifts \eqref{eqn:BCFW.shift} of the type $(-,+)$ (which is valid for $n=12$) involve the fewest number of them. We use for instance the (1,4) shift:
\begin{equation}
\tilde\lambda^{(1)}(z)=\tilde\lambda^{(1)}-z\,\tilde\lambda^{(4)}\,,\qquad \lambda^{(4)}(z)=\lambda^{(4)}+z\,\lambda^{(1)}\,.
\end{equation}
Indeed, it is possible to recover ${\cal A}_{12}$ from only the residues at the following (simple) poles:
\begin{equation}
\ket{2}{4}\,,\;\ket{3}{4}\,,\;\;\Pgen{2}{p_4+p_l}{X}\,,\;\Pgen{3}{p_4+p_l}{X}\quad\textrm{with }l=5,\ldots,12\,,
\label{eqn:poles12}
\end{equation}
which happen, respectively, at the following values of $z$:
\begin{equation}
\underline{z}=-\frac{\ket{2}{4}}{\ket{2}{1}}\,,\;\;\underline{z}=-\frac{\ket{3}{4}}{\ket{3}{1}}\,,\;\;\underline{z}=-\frac{\Pgen{2}{p_4+p_l}{X}}{\ket{2}{1}\bra{4}{X}}\,,\;\;\underline{z}=-\frac{\Pgen{3}{p_4+p_l}{X}}{\ket{3}{1}\bra{4}{X}}\,.
\label{eqn:poles12.z}
\end{equation}
On the one hand, the first two poles coincide with physical ones, and the residue can be computed directly from equation \eqref{eqn:resaln}. We get
\begin{equation}\label{eqn:phys1}\begin{aligned}
\sum_{\textrm{phys}}\textrm{Res}\left[\frac{{\cal A}_{12}(z)}{z}\right]=&-\,\ket{3}{1}\frac{\bra{4}{X}\ket{1}{2}^2}{\ket{2}{4}\ket{1}{4}^2}\textrm{Res}\left[M_{11}^{A}(w),\infty\right]\\
&-\ket{1}{2}\frac{\bra{4}{X}\ket{1}{3}^2}{\ket{3}{4}\ket{1}{4}^2}\textrm{Res}\left[M_{11}^{B}(w),\infty\right]\,,
\end{aligned}\end{equation}
with $M_{11}^{A}$ and $M_{11}^{B}$ being the following eleven-point amplitudes:
\begin{align}
M_{11}^A&=M_{11}\left((1^A)^-,(2^A)^-,3^-,5^+,\ldots,12^+\right)\,,\\
M_{11}^B&=M_{11}\left((1^B)^-,2^-,(3^B)^-,5^+,\ldots,12^+\right)\,,
\end{align}
where notice that particle $4$ has disappeared, and particles 1 and 2, and 1 and 3 respectively, have been deformed as
\begin{align}
P_{1^A}&=\lambda^{(1)}\left(\tilde\lambda^{(1)}+\frac{\ket{2}{4}}{\ket{2}{1}}\tilde\lambda^{(4)}\right)\,, &
P_{2^A}&=\lambda^{(2)}\left(\tilde\lambda^{(2)}+\frac{\ket{1}{4}}{\ket{1}{2}}\tilde\lambda^{(4)}\right)\,,\\
P_{1^B}&=\lambda^{(1)}\left(\tilde\lambda^{(1)}+\frac{\ket{3}{4}}{\ket{3}{1}}\tilde\lambda^{(4)}\right)\,, &
P_{3^B}&=\lambda^{(3)}\left(\tilde\lambda^{(3)}+\frac{\ket{1}{4}}{\ket{1}{3}}\tilde\lambda^{(4)}\right)\,.
\end{align}
One can say that particle 4 has been ``dissolved'' into particles 1 and 2 in $M_{11}^{A}$, and into particles 1 and 3 in $M_{11}^{B}$.

\begin{figure}[h]
\centering
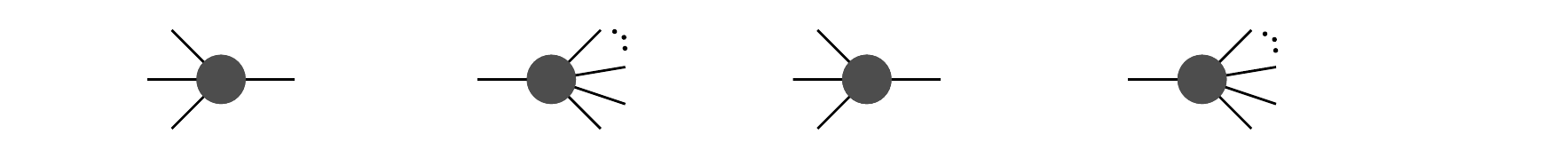
\caption{The only diagrams contributing to the residues at the unphysical poles $\langle 2\vert p_4+p_l\vert X] \text{ (left) and } \langle 3\vert p_4+p_l\vert X]$ (right).}
\label{fig2}
\end{figure}

On the other hand, the residues at the second two types of (unphysical) poles in \eqref{eqn:poles12} can be extracted from just the diagrams of $M_{12}^{\textrm{Ris}}$ where $L^+=\{4,l\}$ and $a=2,3$, which we draw in Figure~\ref{fig2}. 

One just needs to compute the piece proportional to $1/\Pgen{a}{P_{L}}{X}$ of these diagrams. After some simplifications, the result can be compactly written in formula \eqref{eqn:unphys}:
\begin{equation}\label{eqn:unphys1}\begin{aligned}
&\sum_{\textrm{unphys}}\textrm{Res}\left[\frac{{\cal A}_{12}}{z}\right]=-\left(\ket{1}{2}\ket{2}{3}\ket{3}{1}\right)^6\prod_{k=5}^{12}\frac{\bra{k}{X}}{\ket{1}{k}\ket{2}{k}\ket{3}{k}}\\
&\times\sum_{l=5}^{12}\frac{\bra{4}{l}^6}{\bra{4}{X}^2\bra{l}{X}^2}\frac{\ket{4}{l}\bra{4}{l}}{\Pgen{1}{p_4+p_l}{X}\Pgen{2}{p_4+p_l}{X}\Pgen{3}{p_4+p_l}{X}}\frac{\ket{1}{l}\ket{2}{l}\ket{3}{l}}{\bra{l}{X}}\,.
\end{aligned}\end{equation}

Finally, the sum of the two contributions \eqref{eqn:phys1} and \eqref{eqn:unphys1} gives the residue at infinity of the twelve-point amplitude \eqref{eqn:sum.a12} we were looking for. It is quite remarkable that such a complex object admits such a simple expression. Even though \eqref{eqn:phys1} and \eqref{eqn:unphys1} separately are not invariant under permutation of positive labels, their sum is. Therefore the result for the residue at infinity is invariant under permutation of positive (and negative) labels. Both $M_{12}^{\textrm{Ris}}$ and ${\cal A}_{12}$ depend on the reference spinor $X$, but this dependence disappears when we add both terms\footnote{Although this is not obvious from the corresponding analytic expressions, we checked numerically that the final result does not depend on the reference spinor $X$, as well as checking that it agrees with the twelve-graviton amplitude computed via a (more time-consuming) BCFW deformation.}, and we obtain a compact expression for the physical amplitude $M_{12}$.

\section{Discussion on Soft Limits}

The method we proposed to determine the residue at infinity of the twelve-point amplitude from Risager's construction was based on the knowledge of its poles, or equivalently, on the failure of the Risager expansion to provide the correct factorization channels of the physical amplitude. One could think of other possibilities that could also lead to the determination of this residue at infinity.

One such possibility is to look at other singular kinematic limits of the physical amplitude, \textit{e.g}., soft limits. In gravity, the soft factors (which we denote by $S^{\pm}$), defined by
\begin{equation}
\lim_{p_{n+1}\to 0}\frac{M_{n+1}\left(1,\ldots,n,(n+1)^{\pm}\right)}{S^{\pm}_{n+1}\,M_n\left(1,\ldots,n\right)}=1\,,
\end{equation}
are universal\footnote{This means that they depend just on the helicity of the soft graviton, and not on the helicities of the others, \textit{i.e.} they are the same in all sectors. Moreover, they do not get renormalized.}:
\begin{align}
S^-_{n+1}&=\sum_{i=1}^n\frac{\bra{i}{\mu_1}\bra{i}{\mu_2}}{\bra{n+1}{\mu_1}\bra{n+1}{\mu_2}}\frac{\ket{n+1}{i}}{\bra{n+1}{i}}\,,\label{eqn:soft-}\\
S^+_{n+1}&=\sum_{i=1}^n\frac{\ket{i}{\mu_1}\ket{i}{\mu_2}}{\ket{n+1}{\mu_1}\ket{n+1}{\mu_2}}\frac{\bra{n+1}{i}}{\ket{n+1}{i}}\,,\label{eqn:soft+}
\end{align}
where $\mu_1$, $\mu_2$ are arbitrary reference spinors, and the sums above are independent of their choice~\cite{Weinberg:1965nx, Berends:1988zp, Nguyen:2009jk}. We can check if these soft limits (of both negative- and positive-helicity gravitons) are correctly reproduced by the Risager expansion $M_n^\textrm{Ris}$. 

What we find is that soft limits where the momentum of a negative-helicity graviton vanishes produce the correct behavior \eqref{eqn:soft-} on the Risager expansion. However, when the soft graviton has a positive helicity, we obtain the non-trivial behavior:
\begin{equation}
\lim_{p_{l}\to 0}M^{\textrm{Ris}}_{n+1}=\lim_{p_{l}\to 0}\left[S^+_{n+1}M_{n}^{\textrm{Ris}}+\frac{\ket{1}{2}\ket{2}{3}\ket{3}{1}}{\ket{l}{1}\ket{l}{2}\ket{l}{3}}\bra{X}{l}\textrm{Res}\left[M_{n}(w),\infty\right]\right]\,,
\label{eqn:failedsoft}
\end{equation}
where it should be understood that what coincides is the leading order of both sides of the equality, and the graviton $l$ of $M_{n+1}$ is not present in $M_n$. Formula \eqref{eqn:failedsoft} tells us that soft limits are correctly reproduced by $M_n^\textrm{Ris}$ as long as $M_{n-1}^\textrm{Ris}=M_{n-1}$ and $\textrm{Res}\left[M_{n-1}(w),\infty\right]$ vanishes, which happens only for $n<12$. The reason for the failure of the twelve-point Risager expansion to have the right soft limits,
namely the non-vanishing of the residue at infinity of the eleven-point amplitude, 
is exactly the same as the reason why it fails to account for the right residues at the physical poles. Notice the similarities between expressions \eqref{eqn:failedsoft} and \eqref{eqn:resaln}.
 
Given the recent interest in soft factors~\cite{BoucherVeronneau:2011nm, Nandan:2012rk}, and the fact that they play a crucial role in Hodges formula for MHV amplitudes \cite{Hodges:2012ym}, it seems of obvious interest to further explore the possibility of recovering ${\cal A}_n$ from soft limits. Trying to use the formalism of \cite{Benincasa:2011kn} for computing contributions at infinity offers another interesting possibility to recover ${\cal A}_n$. The advantage of this direction would be to have the residue at infinity expressed as a sum of products of MHV amplitudes.

\section*{Acknowledgments}

We would like to thank Freddy Cachazo for numerous conversations, comments on the draft and participation in the early stages of this project. We are also grateful to Bo Feng and David Skinner for useful conversations, and especially to Paolo Benincasa for motivation and comments on the manuscript. E.C. is supported by a Spanish FPU fellowship, and his work is funded in part by MICINN under grants FPA2008-01838 and FPA2011-22594, by the Spanish Consolider-Ingenio 2010 Programme CPAN (CSD2007-00042), by Xunta de Galicia (grant INCITE09 206 121 PR and grant PGIDIT10PXIB206075PR) and by the EU Seventh Framework Programme [FP7-People-2010-IRSES] under grant agreement n$^\circ$269217 (otherwise known as the UNIFY network). E.C. is also grateful to Perimeter Institute for hospitality and to the FRont Of Galician-speaking Scientists for unconditional support.~S.R. is supported in part by the NSERC of Canada and MEDT of Ontario. Research at Perimeter Institute is supported by the Government of Canada through Industry Canada and by the Province of Ontario through the Ministry of Research \& Innovation.

\appendix

\section{Spurious Poles of the Eleven-Point Risager Expansion}
\label{app}

In this Appendix we come back to a technical subtlety of our analysis of Section \ref{ssec:phys}. There we showed analytically how the Risager expansion contains unphysical poles for $n\geq12$. Actually, our analysis naively predicts the presence of an unphysical pole $\bra{l}{X}$ already for the eleven-point Risager expansion, which we know it is not the case as $M_{11}^{\textrm{Ris}}=M_{11}$. Let us see this fact explicitly.

There are three Risager diagrams, that we call $M^{(1,l)},\,M^{(2,l)},\,M^{(3,l)}$, contributing with a $1/\bra{l}{X}$ factor to the eleven-point Risager expansion $M_{11}^{\textrm{Ris}}$. In the expansion \eqref{eqn:Risexp} they are the ones that have a three-particle amplitude with $(1,l,-I)$, $(2,l,-I)$ and $(3,l,-I)$ respectively on the left sub-amplitude. What we have to see is that
\begin{equation}
\lim_{\bra{l}{X}\to0}\bra{l}{X}\left(M^{(1,l)}+M^{(2,l)}+M^{(3,l)}\right)=0\,.
\end{equation}
Let us compute the piece proportional to $1/\bra{l}{X}$ of these diagrams. The left sub-amplitude (leaving $a=1,2,3$ generic) is
\begin{equation}
M_{n_L}=M_3\left(\hat{a}^-,l^+,(-I)^-\right)=\frac{\ket{a}{l}^2}{\bra{a}{X}^2}\bra{l}{X}^6\,,
\label{eqn:miracle11}
\end{equation}
where we are taking into account that
\begin{equation}
\hat{\tilde\lambda}^{(a)}=\frac{\bra{a}{X}}{\bra{l}{X}}\tilde\lambda^{(l)}\,,\quad\lambda^{(I)}=(P_a+P_l)\lvert X]\,,\quad\tilde\lambda^{(I)}=\frac{\tilde\lambda^{(l)}}{\bra{l}{X}}\,.
\end{equation}
The right sub-amplitude is an MHV amplitude, and it looks complicated if we use Mason-Skinner formula \eqref{eqn:MS}. But we just need to keep the leading order in $1/\bra{l}{X}$. Identifying $\hat{b}\equiv 1$, $\hat{c}\equiv n$, $l\equiv n-1$, and using the following result\footnote{Identity \eqref{eqn:identity} was first presented in the context of QED amplitudes (see Section 8.2 of \cite{Mangano:1990by}).} (which is just an elaborated consequence of Schouten identity):
\begin{equation}
\sum_{S_n}\frac{1}{\ket{\alpha}{a_{i_1}}\ket{a_{i_1}}{a_{i_2}}\cdots\ket{a_{i_{n-1}}}{a_{i_n}}\ket{a_{i_n}}{\beta}\ket{\beta}{\alpha}}=\frac{-\ket{\alpha}{\beta}^{n-2}}{\ket{\alpha}{a_1}\cdots\ket{\alpha}{a_n}\ket{a_1}{\beta}\cdots\ket{a_n}{\beta}}\,,
\label{eqn:identity}
\end{equation}
where $S_n$ stands for the permutation group of $n$ elements; we get 
\begin{equation}
M_{n_R}\sim-\frac{\ket{b}{c}^6\Pgen{b}{P_a+P_l}{X}^5\Pgen{c}{P_a+P_l}{X}^5}{\bra{l}{X}^7}\prod_{\substack{k=4\\k\neq l}}^{11}\frac{\bra{k}{l}}{\Pgen{k}{P_a+P_l}{X}\ket{b}{k}\ket{c}{k}}\,.
\label{eqn:MnR}
\end{equation}
Finally, to leading order in $1/\bra{l}{X}$, we obtain
\begin{equation}
M^{(a,l)}\sim-\frac{1}{\bra{l}{X}}\frac{\ket{a}{l}\ket{b}{c}^6\Pgen{b}{P_a+P_l}{X}^5\Pgen{c}{P_a+P_l}{X}^5}{\bra{a}{X}^2\bra{a}{l}}\prod_{\substack{k=4\\k\neq l}}^{11}\frac{\bra{k}{l}}{\Pgen{k}{P_a+P_l}{X}\ket{b}{k}\ket{c}{k}}\,.
\label{eqn:leading11}
\end{equation}
With this we can check if \eqref{eqn:miracle11} is satisfied. In the limit $\bra{l}{X}\to0$ we can put, up to unimportant scaling factors, $\tilde\lambda^{(l)}=X$. Then we have
\begin{multline}
\lim_{\bra{l}{X}\to0}\bra{l}{X}\left(M^{(1,l)}+M^{(2,l)}+M^{(3,l)}\right)=\left(\ket{1}{2}\ket{2}{3}\ket{3}{1}\right)^5\\
\times\prod_{\substack{k=4\\k\neq l}}^{11}\frac{\bra{k}{l}}{\ket{1}{k}\ket{2}{k}\ket{3}{k}}\left(\ket{1}{l}\ket{2}{3}+\ket{2}{l}\ket{3}{1}+\ket{3}{l}\ket{1}{2}\right)=0\,,
\end{multline}
where we just used Schouten identity in the last line. This shows that the pole $\bra{l}{X}$, appearing in three of the Risager diagrams, is spurious as it cancels when summing over them. Notice also that this cancellation is only possible when $n=11$, where $\Pgen{b}{P_a+P_l}{X}^5$ and $\Pgen{c}{P_a+P_l}{X}^5$ in~\eqref{eqn:MnR} come exactly with the power five.

\end{document}